\documentclass[intlimits,twoside,a4paper]{article}
\usepackage{amsmath,amssymb}
\usepackage{graphicx}
\usepackage{color}
\usepackage[T2A]{fontenc}
\usepackage[cp1251]{inputenc}

\usepackage[]{cmpj2}

\newtheorem{theorem}{Theorem}

\newtheorem{example}[theorem]{Example}

\issue{2017}{20}{1}{13001}
\doinumber{10.5488/CMP.20.13001}

\title[Investigating inequality: a Langevin approach]{Investigating inequality: a Langevin approach}
\author[I. Eliazar]{I. Eliazar\thanks{E-mail: iddo.eliazar@intel.com, eliazar@post.tau.ac.il.}}
\address{Smart Device Innovation Science Team, New Devices Group, Intel Corporation,
Yakum, Israel}

\authorcopyright{I. Eliazar, 2017}
\date{Received November 26, 2016}

\begin{document}

\maketitle

\begin{abstract}
Inequality indices are quantitative scores that gauge the divergence of
wealth distributions in human societies from the ``ground state'' of pure
communism. While inequality indices were devised for socioeconomic
applications, they are effectively applicable in the context of general
non-negative size distributions such as count, length, area, volume, mass,
energy, and duration. Inequality indices are commonly based on the notion of
Lorenz curves, which implicitly assume the existence of finite means.
Consequently, Lorenz-based inequality indices are excluded from the realm of
infinite-mean size distributions. In this paper we present an inequality
index that is based on an altogether alternative Langevin approach. The
Langevin-based inequality index is introduced, explored, and applied to a
wide range of non-negative size distributions with both finite and infinite
means.

\pacs 02.50.-r, 89.65.-s
\end{abstract}

\section{\label{1}Introduction}

This paper is written in honor of the 60$^{\text{th}}$ birthday of Professor
Yurij Holovatch. The scientific interests of Professor Holovatch span from
statistical physics to econophysics and sociophysics. In this paper we'll
try to weave these interests together via a topic that draws considerable
attention among the aforementioned physics audiences: \emph{inequality} \cite{CYC,CCC,CCCC,SC,Buc,Cha}.

Following the pioneering work of Vilfredo Pareto on the distribution of
wealth in human societies \cite{Par}, the study of \emph{socioeconomic
inequality} became a major topic in economics and in the social sciences, as
well as a major topic of wide public debate \cite{ABD,Nec,DA,Mil1,Fre,Mil2,Sut,Sti1,Dea,Sci,CP,Atk1,Bou2,Sti2,Boi,Pik,Leo,Mil3,Lin}.
Scientists devised special metrics termed \emph{inequality indices} to gauge
socioeconomic inequality \cite{Cou,BL,HN,Cow}, the most notable of which
being the \emph{Gini index} \cite{Gin1,Gin2,Yit,YS} and the \emph{Pietra index}
\cite{Pie,Hoo,Sar,ES2,Fro,SJ}. Inequality indices measure the divergence of wealth
distributions from the ``ground state'' of pure communism, and score this
divergence. From an abstract mathematical perspective inequality indices can
be applied in the context of general distributions of non-negative sizes ---
e.g., count, length, area, volume, mass, energy, duration, etc. --- and serve,
in this general context, as gauges of statistical heterogeneity \cite{StaEve,HarInq}.

Inequality indices are based on the notion of \emph{Lorenz curves} \cite{Lor,Gas,Gio,Arn1,Cho}, which represent wealth distributions --- and, in general,
distributions of non-negative sizes --- in a universally calibrated
statistical method. More specifically, inequality indices are based either
on geometric divergence \cite{SocG1,SocG2} or on entropic divergence
\cite{InvEq,RenEq}, from the ``ground state'' of pure communism, in the
space of Lorenz curves. Lorenz curves implicitly require a \emph{finite mean}
of their input distributions. Consequently, inequality indices are
applicable only in the case of non-negative size distributions with finite
means. However, non-negative size distributions with \emph{infinite means}
are prevalent across the sciences and are often of prime importance \cite{BG,JW,ST,MK,KS,New,Red,CSN,Nol,EK}.

In this paper we propose a new ``Langevin approach'' to inequality. The
approach is based on the \emph{Langevin equation} \cite{Lan,CK}, a
foundational notion in statistical physics that stands at the core of
diffusion processes and stochastic dynamics \cite{Gar,VK}. Following
up on the ``topography of chance'' that emanates from the Langevin equation in
the setting of U-shaped potential landscapes \cite{TopCha}, we introduce and
explore a novel scenario-based inequality index that is applicable for a
wide range of non-negative size distributions with both finite and infinite
means.

In a sense this paper ``closes a circle''. Indeed, in \cite{StaEve,HarInq} it was demonstrated how the socioeconomic methods of Lorenz curves
and inequality indices can be ``imported'' to the physical sciences, and serve
there to quantify the statistical heterogeneity of general non-negative size
distributions with finite means. This paper moves in the opposite direction,
and demonstrates how a Langevin-based approach can effectively replace the
Lorenz-based approach of gauging inequality. This is yet another example of
the importance of the trans-disciplinary flow of ideas between different
fields of science.

The paper is organized as follows. We begin with a brief review of the
Langevin equation and its associated Gibbs density (section \ref{2}),
followed by a brief review of the corresponding geometric Langevin equation
and its associated geometric Gibbs density (section \ref{3}). After
describing the case of underlying U-shaped potential landscapes (section \ref%
{4}), we introduce a scenario-based equality index (sections \ref{5} and~\ref%
{6}). We then describe the general application of the scenario-based
equality index as a gauge of inequality (section \ref{7}), and conclude with
illustrative examples of the application (section \ref{8}).

\section{\label{2}Gibbs density}

A key pillar of statistical physics is the notion of \emph{entropy
maximization}: characterizing the most random configuration within a class
of configurations that is determined by a given set of constraints \cite{Jay,Kap}. Specifically, consider the class of probability density
functions over the real line, $\phi \left( x\right) $ ($-\infty <x<\infty $%
), that are subject to the following moment constraint: $\int_{-\infty
}^{\infty }V\left( x\right) \phi \left( x\right) \rd x=v$, where $V\left(
x\right) $ is an general moment function, and where $v$ is an admissible
moment value. Within this class of densities the unique density that
maximizes the Boltzmann-Gibbs-Shannon entropy \cite{Bol,Gib,Sha} is the
following \emph{Gibbs density}:%
\begin{equation}
\phi \left( x\right) =c\exp \left[ -\frac{1}{\tau }V\left( x\right) \right],  \label{201}
\end{equation}%
where $c$ is a positive normalization constant, and where $\tau $ is a
positive ``temperature'' parameter that is set to meet the moment constraint.

Yet another key pillar of statistical physics is the \emph{Langevin equation}%
: a stochastic differential equation that describes motion on a potential
landscape that is subject to random white-noise fluctuations \cite{Lan,CK}. Specifically, consider a potential landscape over the real line ($%
-\infty <x<\infty $) that is given by the function $V\left( x\right) $, set $%
F\left( x\right) =-V^{\prime }\left( x\right) $ to be the corresponding
``force function'', and denote by $X\left( t\right) $ the real-valued position
of the motion at time $t$. The stochastic dynamics of the motion are governed by
the following Langevin equation:%
\begin{equation}
\dot{X}\left( t\right) =F\left[ X\left( t\right) \right] +\sigma \dot{W}%
\left( t\right), \label{202}
\end{equation}%
where $\sigma $ is a positive ``volatility'' parameter, and where $\dot{W}%
\left( t\right) $ is white noise \cite{Gar,VK}.\footnote{White noise $\dot{W}\left( t\right) $ is a stochastic process that is
characterized by the following pair of properties \cite{Gar,VK}: (i)
its integral over an arbitrary time interval, $\int_{a}^{b}\dot{W}\left(
t\right) \rd t$ ($a<b$), is a random variable that is governed by a \emph{%
normal law} with mean zero and with variance $b-a$; and (ii) its integrals
over disjoint time intervals are independent random variables.}

There is a profound connection between these two seemingly unrelated
pillars, the notion of entropy maximization and the Langevin equation.
Indeed, the \emph{stationary law} of the motion whose stochastic dynamics
are governed by the Langevin equation~(\ref{202}) is given by the Gibbs
density of equation~(\ref{201}) with temperature $\tau =\sigma ^{2}/2$. Namely,
consider the steady-state position of motion: $X\left( \infty \right)
=\lim_{t\rightarrow \infty }X\left( t\right) $, the limit being in law.
Then: the law of the random variable $X\left( \infty \right) $ is governed
by the Gibbs density of equation~(\ref{201}) with $\tau =\sigma ^{2}/2$.

\section{\label{3}Geometric Gibbs density}

The Langevin equation~(\ref{202}) describes \emph{additive} evolution. In
many processes, especially in economics and finance, evolution is \emph{%
multiplicative} rather than additive \cite{LSS,RL,Hul,GeoLan,MS}. The
transformation from additive evolution to multiplicative evolution is
facilitated by exponentiation. Specifically, in the context of the Langevin
equation~(\ref{202}): consider $X\left( t\right) $ to be the log-scale
position at time $t$, and shift to the position $Y\left( t\right) =\exp %
\left[ X\left( t\right) \right] $. Applying this exponential transformation,
Ito's formula \cite{Ito,IM}\ implies that the stochastic dynamics of
the exponentiated motion are governed by the following \emph{geometric
Langevin equation}:%
\begin{equation}
\frac{\dot{Y}\left( t\right) }{Y\left( t\right) }=G\left[ Y\left( t\right) %
\right] +\sigma \dot{W}\left( t\right) ,  \label{301}
\end{equation}%
where the ``geometric force function'' is given by $G( y) =\tau +F%
[\ln ( y)] $ ($0<y<\infty $), with $\tau =\sigma
^{2}/2$. Comparing equations~(\ref{202}) and (\ref{301}) it is indeed evident
that the former describes additive evolution, whereas the latter describes
multiplicative evolution.

In turn, the \emph{stationary law} of the motion whose stochastic dynamics
are governed by the geometric Langevin equation~(\ref{301}) is given by the
following \emph{geometric Gibbs density}:%
\begin{equation}
\psi ( y) =c\exp \left\{ -\frac{1}{\tau }V\left[ \ln (y) \right] \right\} \frac{1}{y}  \label{302}
\end{equation}%
($0<y<\infty $), with temperature $\tau =\sigma ^{2}/2$. Namely, consider
the steady-state position of the exponentiated motion: $Y\left( \infty
\right) =\lim_{t\rightarrow \infty }Y\left( t\right) $, the limit being in
law. Then: the law of the random variable $Y\left( \infty \right) $ is
governed by the geometric Gibbs density of equation~(\ref{302}) with $\tau
=\sigma ^{2}/2$.

\section{\label{4}Shapes}

One of the most fundamental potential landscapes, in the context of the
Langevin equation, is that of a \emph{U-shaped} ``potential well'' \cite%
{TopCha}: $V\left( x\right) $ being a convex function with a unique global
minimum. The corresponding force function admits the following shape: $%
F\left( x\right) $ is a continuous function that de\-creases mo\-no\-to\-nous\-ly
from a positive limit $F\left( -\infty \right) :=\lim_{x\rightarrow -\infty
}F\left( x\right) >0$ to a negative limit $F\left( \infty \right)
:=\lim_{x\rightarrow \infty }F\left( x\right) <0$. In what follows we
consider this U-shaped potential landscape to hold, and denote by $%
F^{-1}\left( \cdot \right) $ the inverse of the corresponding monotonously
decreasing force function.

The stationary Gibbs density $\phi \left( x\right) $ of the Langevin
equation~(\ref{202}) admits the following unimodal shape: it is monotonously
increasing on the half-line $-\infty <x<F^{-1}\left( 0\right) $, it attains
its unique global maximum at the point $F^{-1}\left( 0\right) $, and it is
monotonously decreasing on the half-line $F^{-1}\left( 0\right) <x<\infty $.
Hence, the \emph{mode} $x_{\ast }$ of the stationary Gibbs density $\phi
\left( x\right) $ is the point at which the force function $F\left( x\right)
$ crosses the level zero: $x_{\ast }=F^{-1}\left( 0\right) $. The core of
the Langevin equation~(\ref{202}) is the ordinary differential equation $%
\dot{x}\left( t\right) =F\left[ x\left( t\right) \right] $. The \emph{%
stationary solution} of this ordinary differential equation, $x\left( \infty
\right) =\lim_{t\rightarrow \infty }x\left( t\right) $, is its fixed point
--- i.e., the point at which the force function $F\left( x\right) $ crosses
the level zero: $x\left( \infty \right) =F^{-1}\left( 0\right) $. Thus, we
observe that the mode of the stationary Gibbs density $\phi \left( x\right) $
coincides with the stationary solution of the ordinary differential equation
$\dot{x}\left( t\right) =F\left[ x\left( t\right) \right] $, i.e.: $x_{\ast
}=x\left( \infty \right) $. We shall now argue that the transition from the
additive evolution of the Langevin equation~(\ref{202}) to the
multiplicative evolution of the geometric Langevin equation~(\ref{301})
breaks this coincidence.

If $F\left( -\infty \right) \leqslant \tau $, then the stationary geometric Gibbs
density $\psi ( y) $ of the geometric Langevin equation~(\ref{301}%
) is monotonously decreasing on the positive half-line $0<y<\infty $. On the
other hand, if $F\left( -\infty \right) >\tau $, then the stationary
geometric Gibbs density $\psi ( y) $ admits the following
unimodal shape: it is monoto\-nously increasing on the interval $0<y<\exp
[F^{-1}\left( \tau \right) ]$, it attains its unique global maximum at the
point \linebreak $\exp [F^{-1}\left( \tau \right) ]$, and it is monotonously decreasing on
the half-line $\exp [F^{-1}\left( \tau \right) ]<y<\infty $. Hence, the
\emph{mode} of the stationary geometric Gibbs density $\psi ( y) $
is given by%
\begin{equation}
y_{\ast }=\left\{
\begin{array}{lll}
0, &  \text{if }&F\left( -\infty \right) \leqslant \tau, \\
\exp \left[F^{-1}\left( \tau \right) \right], & \text{if } &F\left( -\infty \right)
>\tau.%
\end{array}%
\right.  \label{401}
\end{equation}

The core of the geometric Langevin equation~(\ref{301}) is the ordinary
differential equation $\dot{y}\left( t\right) =G[ y\left( t\right)] $. If $F\left( \infty \right) <-\tau $, then the stationary solution
of this ordinary differential equation, $y\left( \infty \right)
=\lim_{t\rightarrow \infty }y\left( t\right) $, is its fixed point --- i.e.,
the point at which the geometric force function $G( y) $ crosses
the level zero: $y\left( \infty \right) =G^{-1}\left( 0\right) $; note that $%
G^{-1}\left( 0\right) =\exp [F^{-1}\left( -\tau \right) ]$. On the other
hand, if $F\left( \infty \right) \geqslant -\tau $, then the stationary solution
diverges: $y\left( \infty \right) =\infty $. Hence, the \emph{stationary
solution} of the ordinary differential equation $\dot{y}\left( t\right) =G[ y\left( t\right) ] $ is given by
\begin{equation}
y\left( \infty \right) =\left\{
\begin{array}{lll}
\exp \left[F^{-1}\left( -\tau \right) \right], &  \text{if }&F\left( \infty \right)
<-\tau , \\
\infty,  & \text{if } &F\left( \infty \right) \geqslant -\tau. %
\end{array}%
\right.   \label{402}
\end{equation}

It is evident from equations~(\ref{401}) and (\ref{402}) that the mode of the
stationary geometric Gibbs density $\psi ( y) $ is always \emph{%
smaller} than the stationary solution of the ordinary differential equation $%
\dot{y}\left( t\right) =G[ y\left( t\right) ] $, i.e.: $y_{\ast
}<y\left( \infty \right) $. Thus, as claimed above, the transition from the
additive evolution of the Langevin equation~(\ref{202}) to the
multiplicative evolution of the geometric Langevin equation~(\ref{301})
indeed breaks the coincidence of the mode and the stationary solution: we
shift from\emph{\ }$x_{\ast }=x\left( \infty \right) $ to $y_{\ast }<y\left(
\infty \right) $.

\section{\label{5}Equality: Langevin representation}

An \emph{equality index} $\mathcal{E}$ is a gauge of positive-valued random
variables that scores their inherent equality, i.e., their inherent
statistical homogeneity \cite{InvEq,RenEq}. Equality indices take
values in the unit interval, $0\leqslant \mathcal{E}\leqslant 1$, yield the maximal
unit score only in the case of constant random variables, and are
scale-invariant (we shall elaborate on the scale-invariance property in
section \ref{7} below).

In the context of the Langevin equation~(\ref{202}) and of the geometric
Langevin equation~(\ref{301}), the respective steady-state random variables $%
X\left( \infty \right) $ and $Y\left( \infty \right) $ are constant only
when no randomness is present. The absence of randomness is attained by
setting the volatility, and hence the temperature, to vanish: $\tau =\sigma
^{2}/2=0$. Using the notation of the previous section, zero temperature
implies that $X\left( \infty \right) =x\left( \infty \right) $ and $Y\left(
\infty \right) =\exp \left[ x\left( \infty \right) \right] $. Focusing on
the geometric Langevin equation~(\ref{301}), the notation of the previous
section further implies that%
\begin{equation}
0\leqslant y_{\ast }<\exp \left[ F^{-1}\left( 0\right) \right] <y\left( \infty
\right) \leqslant \infty,  \label{500}
\end{equation}%
where:

\begin{enumerate}
\item[$\bullet $] $y_{\ast }$ is the mode of the stationary geometric Gibbs
density $\psi( y) $, and is given by equation~(\ref{401}).

\item[$\bullet $] $\exp [ F^{-1}\left( 0\right)] $ is the
constant value of the steady-state random variable $Y\left( \infty \right) $%
, attained at zero temperature.

\item[$\bullet $] $y\left( \infty \right) $ is the stationary solution of
the ordinary differential equation $\dot{y}\left( t\right) =G[ y\left(
t\right)] $, and is given by equation~(\ref{402}).
\end{enumerate}

Equation~(\ref{500}) holds for any positive temperature, $\tau >0$. When the
temperature vanishes, $\tau =0$, the strict inequalities of equation~(\ref{500})
become equalities: $y_{\ast }=\exp [ F^{-1}\left( 0\right)]
=y\left( \infty \right) $. Thus, we can use the divergence of the three
terms of equation~(\ref{500}) from each other in order to gauge the inherent
equality of the steady-state random variable $Y\left( \infty \right) $.
Indeed, each of the three following ratios can serve as an equality index:

\begin{enumerate}
\item[$\bullet $] The ``left-hand ratio''\ $\mathcal{R}_{l}:=y_{\ast }/\exp [
F^{-1}\left( 0\right) ] $, which vanishes when $F\left( -\infty
\right) \leqslant \tau $.

\item[$\bullet $] The ``right-hand ratio''\ $\mathcal{R}_{r}:=\exp [
F^{-1}\left( 0\right)] /y\left( \infty \right) $, which vanishes when
$F\left( \infty \right) \geqslant -\tau $.

\item[$\bullet $] The ``overall ratio''\ $\mathcal{R}:=y_{\ast }/y\left(
\infty \right) $, which vanishes when either $F\left( -\infty \right) \leqslant
\tau $ or $F\left( \infty \right) \geqslant -\tau $.
\end{enumerate}

We combine these three ratios together into a joint \emph{scenario-based
equality index} $\mathcal{E}$ that is presented in table~\ref{tab1}. Equations~(\ref{401})
and (\ref{402}) imply the ratio values that are specified in table~\ref{tab1} for
each scenario.

\begin{table}[!t]
\caption{The scenario-based equality index $\mathcal{E}$. The rows
indicate whether the force-function limit $F\left( -\infty \right)
:=\lim_{x\rightarrow -\infty }F\left( x\right) $ is greater than the
temperature $\tau $, or not. The columns indicate whether the force-function
limit $F\left( \infty \right) :=\lim_{x\rightarrow \infty }F\left( x\right) $
is smaller than minus the temperature $-\tau $, or not. The two rows and
the two columns determine four different scenarios, and for each scenario
the equality index $\mathcal{E}$ is specified: the ratio $\mathcal{R}$, the
ratio $\mathcal{R}_{l}$, the ratio $\mathcal{R}_{r}$, or zero.}
\label{tab1}
\vspace{2ex}
\begin{center}
\begin{tabular}{|c|c|c|}
\hline\hline
& $F\left( \infty \right) <-\tau $ & $F\left( \infty \right) \geqslant -\tau$ \\
\hline
$F\left( -\infty \right) >\tau $ & $\mathcal{R}=\exp \left[F^{-1}\left( \tau \right) -F^{-1}\left( -\tau \right) \right] $ & $\mathcal{R}_{l}=\exp \left[F^{-1}\left( \tau \right) -F^{-1}\left( 0\right) \right] $ \\
\hline
$F\left( -\infty \right) \leqslant \tau $ & $\mathcal{R}_{r}=\exp \left[F^{-1}\left( 0\right) -F^{-1}\left( -\tau \right) \right] $ & $0 $ \\
\hline\hline
\end{tabular}
\end{center}
\end{table}
\vspace{-2mm}

\section{\label{6}Equality: Density representation}

In the previous section we introduced the scenario-based equality index $%
\mathcal{E}$, and represented it in terms of the underlying
Langevin-equation structure: the force function $F\left( x\right) $ and the
temperature $\tau =\sigma ^{2}/2$. This representation is applicable when
the underlying Langevin-equation structure is known. However, if this
underlying structure is unobservable, and only the stationary geometric
Gibbs density $\psi ( y) $ is observable, an alternative
representation is required. To that end, we introduce the function
\begin{equation}
\varphi ( y) =y\frac{\psi ^{\prime }( y) }{\psi (
y) }  \label{601}
\end{equation}%
($0<y<\infty $), and note that in terms of the function $\varphi (
y) $ the Langevin force function is given by:
\begin{equation}
\frac{1}{\tau }F\left( x\right) =1+\varphi \left[ \exp \left( x\right) %
\right]  \label{602}
\end{equation}%
($-\infty <x<\infty $). Consequently, the Langevin-based representation of
the scenario-based equality index $\mathcal{E}$ --- which was specified in
table~\ref{tab1} --- shifts to a density-based representation. Namely, table~\ref{tab2}
specifies the representation of the scenario-based equality index $\mathcal{E%
}$ in terms of the function $\varphi ( y) $ of equation~(\ref{601}%
).

\begin{table}[!h]
\caption{The scenario-based equality index $\mathcal{E}$,
represented in terms of the function $\varphi ( y) $ of equation~(\ref%
{601}). The rows indicate whether the limit $\varphi \left( 0\right)
:=\lim_{y\rightarrow 0}\varphi ( y) $ is positive, or not. The
columns indicate whether the limit $\varphi \left( \infty \right)
:=\lim_{y\rightarrow \infty }\varphi ( y) $ is smaller than $-2$,
or not. The two rows and the two columns determine the four different
scenarios, and for each scenario the equality index $\mathcal{E}$ is
specified: the ratio $\mathcal{R}$, the ratio $\mathcal{R}_{l}$, the ratio $%
\mathcal{R}_{r}$, or zero.}
\label{tab2}
\vspace{2ex}
\begin{center}
\begin{tabular}{|c|c|c|}
\hline\hline
& $\varphi \left( \infty \right) <-2 $ & $\varphi \left( \infty \right) \geqslant -2 $ \\
\hline
$\varphi \left( 0\right) >0 $ & $\mathcal{R}=\frac{\varphi ^{-1}\left( 0\right) }{\varphi ^{-1}\left(
-2\right) }$ & $\mathcal{R}_{l}=\frac{\varphi ^{-1}\left( 0\right) }{\varphi ^{-1}\left(
-1\right) }$ \\
\hline
$\varphi \left( 0\right) \leqslant 0 $ & $\mathcal{R}_{r}=\frac{\varphi ^{-1}\left( -1\right) }{\varphi ^{-1}\left(
-2\right) }$ & $0$ \\
 \hline\hline
\end{tabular}
\end{center}
\end{table}
\vspace{-7mm}
\section{\label{7}Application}

Table~\ref{tab2} facilitates the wide application of the scenario-based equality
index $\mathcal{E}$. Indeed, assume that we are given a positive-valued
random variable $Y$, with probability density function $\psi ( y)
$ ($0<y<\infty $), to which we want to apply the scenario-based equality
index $\mathcal{E}$. The first step is to check whether the underlying Langevin
setting --- with a U-shaped potential landscape --- applies to the given
random variable $Y$. Namely: is the underlying force $F\left( x\right) $ a
continuous function that decreases monotonously from a positive limit $%
F\left( -\infty \right) :=\lim_{x\rightarrow -\infty }F\left( x\right) >0$
to a negative limit $F\left( \infty \right) :=\lim_{x\rightarrow \infty
}F\left( x\right) <0$? Equation~(\ref{602}) implies that this question, in terms
of the function $\varphi ( y) $ of equation~(\ref{601}), translates to
the following question: is $\varphi ( y) $ a continuous function
that decreases monotonously from the limit $\varphi \left( 0\right)
:=\lim_{y\rightarrow 0}\varphi ( y) >-1$ to the limit $\varphi
\left( \infty \right) :=\lim_{y\rightarrow \infty }\varphi ( y)
<-1$? If the answer to the latter question is affirmative, then the
scenario-based equality index $\mathcal{E}$ is applicable indeed, and we may
proceed to the second step: computing this index, via the function $\varphi
( y) $ of equation~(\ref{601}), according to table~\ref{tab2}.

At the beginning of section \ref{5} we noted that equality indices are
required to be \emph{scale invariant}. Equality indices were originally
devised in order to gauge the socioeconomic equality of wealth distributions
in human societies. Evidently, an equality score should be invariant with
respect to the specific currency via which wealth is measured. Indeed, the
same human society has the same degree of socioeconomic equality --- no
matter in which currency the wealth of the society members is measured. In
the context of wealth distributions, the random variable $Y$ represents the
wealth of a randomly sampled society member, and a change of currency is
manifested by the following \emph{scale transformation}: $Y\rightarrow  cY$,
where $c$ is a positive constant. In turn, equality indices should be
invariant with respect to this scale transformation. The scenario-based
equality index $\mathcal{E}$ indeed satisfies the scale-invariance property:
table~\ref{tab2} that corresponds to the scaled random variable $cY$ is identical
to table~\ref{tab2} that corresponds to the random variable~$Y$.

Another transformation we address is \emph{reciprocation}: $Y\rightarrow 1/Y$.
On the log-scale, $X=\ln \left( Y\right) $, reciprocation manifests
mirroring: $X\rightarrow -X$. Reciprocation switches between the roles of the
small ``poor''\ values and the large ``rich''\ values. The effect of
reciprocation on the scenario-based equality index $\mathcal{E}$ is rather
``elegant'': table~\ref{tab2} that corresponds to the reciprocated random variable $%
1/Y$ is given by the \emph{transposition} of table~\ref{tab2} that corresponds to
the random variable $Y$.

Yet another transformation we address is the \emph{power transformation}: $%
Y\rightarrow Y^{p}$, where $p$ is a positive power. The power transformation is
prevalent across the sciences, and the scenario-based equality index $%
\mathcal{E}$ that corresponds to the power random variable $Y^{p}$ is
presented in table~\ref{tab3}. Note that for the power $p=1$ table~\ref{tab3} indeed yields
back table~\ref{tab2}.

\begin{table}[!h]
\caption{The scenario-based equality index $\mathcal{E}$ of the
power random variable $Y^{p}$, represented in terms of the function $\varphi
( y) $ of equation~(\ref{601}). The rows indicate whether the limit $%
\varphi \left( 0\right) :=\lim_{y\rightarrow 0}\varphi ( y) $ is
greater than $p-1$, or not. The columns indicate whether the limit $\varphi
\left( \infty \right) :=\lim_{y\rightarrow \infty }\varphi ( y) $
is smaller than $-p-1$, or not. The two rows and the two columns
determine the four different scenarios, and for each scenario the equality
index $\mathcal{E}$ is specified: the ratio $\mathcal{R}$, the ratio $%
\mathcal{R}_{l}$, the ratio $\mathcal{R}_{r}$, or zero.}
\label{tab3}
\vspace{2ex}
\begin{center}
\begin{tabular}{|c|c|c|}
\hline\hline
& $\varphi \left( \infty \right) <-p-1 $ & $\varphi \left( \infty \right) \geqslant -p-1 $ \\
\hline
$\varphi \left( 0\right) >p-1 $ & $\mathcal{R}=\left[ \frac{\varphi ^{-1}\left( p-1\right) }{\varphi
^{-1}\left( -p-1\right) }\right] ^{p}$ & $\mathcal{R}_{l}=\left[ \frac{\varphi ^{-1}\left( p-1\right) }{\varphi
^{-1}\left( -1\right) }\right] ^{p}$ \\
\hline
$\varphi \left( 0\right) \leqslant p-1 $ & $\mathcal{R}_{r}=\left[ \frac{\varphi ^{-1}\left( -1\right) }{\varphi
^{-1}\left( -p-1\right) }\right] ^{p}$ & $0$ \\
\hline\hline
\end{tabular}
\end{center}
\end{table}

In sections \ref{5} and \ref{6}, as well as in this section, we focused on
the notion of \emph{equality}. In fact, as noted in the introduction, it is
more common to address \emph{inequality }rather than equality. An \emph{%
inequality index} $\mathcal{I}$ is a gauge of positive-valued random
variables that scores their inherent inequality, i.e., their inherent
statistical heterogeneity \cite{Cou,BL,HN,Cow}. Inequality indices take
values in the unit interval, $0\leqslant \mathcal{I}\leqslant 1$, yield the minimal
zero score only in the case of constant random variables, and are
scale-invariant (as described above). The shift between gauging equality and
inequality is straightforward. Indeed, every equality index $\mathcal{E}$
has a coupled inequality index $\mathcal{I}$ (and vice-versa), and the
coupling is given by: $\mathcal{E}+\mathcal{I}=1$. Hence, all the results
regarding the scenario-based equality index $\mathcal{E}$ can be shifted to
the coupled scenario-based inequality index, $\mathcal{I}=1-\mathcal{E}$.

\section{\label{8}Examples}

To demonstrate the application of the scenario-based equality index $%
\mathcal{E}$, we present in this section four illustrative examples. In all
the examples we are given a positive-valued random variable $Y$ with
probability density function $\psi ( y) $ ($0<y<\infty $). In all
the densities, the parameter $c$ denotes a normalization constant, and in the
densities of examples \ref{E2}--\ref{E4} the parameters $\alpha $ and $\beta $
denote positive exponents.

\begin{example}
\label{E1}
\begin{equation}
\psi ( y) =c\frac{1}{y}\exp \left\{ -\frac{\left[ \ln (
y) -m\right] ^{2}}{2v}\right\} .  \label{710}
\end{equation}%
\emph{This density characterizes the }lognormal law\emph{\ with a real
log-scale mean }$m$\emph{\ and a positive log-scale variance }$v$\emph{\
\cite{AB,CS,JKB,LSA,Mit}. Applying table~\ref{tab2} we obtain that}%
\begin{equation}
\mathcal{E}=\mathcal{R}=\exp \left( -2v\right) .  \label{711}
\end{equation}%
\emph{Namely, in this example only the scenario }$\mathcal{E=R}$\emph{\
takes place.}
\end{example}

\begin{example}
\label{E2}
\begin{equation}
\psi ( y) =c\exp \left( -y^{\alpha }\right) y^{\beta -1}.
\label{720}
\end{equation}%
\emph{This density generalizes the }gamma law\emph{\ (}$\alpha =1$\emph{)
\cite{Fel} and the }Weibull extreme-value law\emph{\ (}$\alpha =\beta $\emph{%
) \cite{Wei1,Wei2,Rin,GeoWei}. Applying table~\ref{tab2} we obtain that}%
\begin{equation}
\mathcal{E}=\left\{
\begin{array}{ccc}
\vspace{2mm}
\mathcal{R}=\left( \frac{\beta -1}{\beta +1}\right) ^{1/\alpha }, &  \emph{if} & \beta >1, \\
\mathcal{R}_{r}=\left( \frac{\beta }{\beta +1}\right) ^{1/\alpha }, &  \emph{if} & \beta \leqslant 1.
\end{array}%
\right.   \label{721}
\end{equation}%
\emph{Namely, in this example only the scenarios }$\mathcal{E}=\mathcal{R}$
\emph{and} $\mathcal{E}=\mathcal{R}_{r}$ \emph{take place.}
\end{example}

\begin{example}
\label{E3}
\begin{equation}
\psi ( y) =c\exp \left( -y^{-\alpha }\right) y^{-\beta -1}\text{ .%
}  \label{730}
\end{equation}%
\emph{This density generalizes the }inverse-gamma law\emph{\ (}$\alpha =1$%
\emph{) \cite{Ber,Bols,Lee} and the }Fr\'{e}chet extreme-value law\emph{\ (%
}$\alpha =\beta $\emph{) \cite{Gal,KN,RT}. Applying table~\ref{tab2} we obtain
that}%
\begin{equation}
\mathcal{E}=\left\{
\begin{array}{ccc}
\vspace{2mm}
\mathcal{R}=\left( \frac{\beta -1}{\beta +1}\right) ^{1/\alpha }, &  \emph{if} & \beta >1, \\
\mathcal{R}_{l}=\left( \frac{\beta }{\beta +1}\right) ^{1/\alpha }, &  \emph{if} &
\beta \leqslant 1.%
\end{array}%
\right.   \label{731}
\end{equation}%
\emph{Namely, in this example only the scenarios} $\mathcal{E}=\mathcal{R}$
\emph{and} $\mathcal{E}=\mathcal{R}_{l}$ \emph{take place}.
\end{example}

\begin{example}
\label{E4}
\begin{equation}
\psi ( y) =c\frac{y^{\alpha -1}}{\left( 1+y\right) ^{\alpha
+\beta }}\,.  \label{740}
\end{equation}%
\emph{This density characterizes the }beta-prime law\emph{, also known as
the }Fisher-Snedecor law\emph{\ \cite{IN,Sff,Hoc}. The scena\-rio-based
equality index }$\mathcal{E}$\emph{\ of this law is given in table~\ref{tab4}.}
\end{example}

\begin{table}[!h]
\caption{The scenario-based equality index $\mathcal{E}$
corresponding to the beta-prime density of equation~(\ref{740}). The rows
indicate whether the exponent $\alpha $ is greater than $1$, or not. The
columns indicate whether the exponent $\beta $ is greater than $1$, or not.
The two rows and the two columns determine the four different scenarios, and
for each scenario the equality index $\mathcal{E}$ is specified: the ratio $%
\mathcal{R}$, the ratio $\mathcal{R}_{l}$, the ratio $\mathcal{R}_{r}$, or
zero.}
\label{tab4}
\vspace{2ex}
\begin{center}
\begin{tabular}{|c|c|c|}
\hline\hline
& $\beta >1 $ & $\beta \leqslant 1 $ \\
\hline
$\alpha >1 $ & $\mathcal{R}=\frac{\alpha -1}{\alpha +1}\frac{\beta -1}{\beta +1}$ & $\mathcal{R}_{l}=\frac{\alpha -1}{\alpha }\frac{\beta }{\beta +1}$ \\
\hline
$\alpha \leqslant 1 $ & $\mathcal{R}_{r}=\frac{\alpha }{\alpha +1}\frac{\beta -1}{\beta }$ & $0$ \\
\hline\hline
\end{tabular}
\end{center}
\end{table}

Several remarks regarding examples \ref{E2}--\ref{E4}. First, the similarity
between examples \ref{E2} and \ref{E3} is not coincidental: indeed, the
random variables of these examples are related by reciprocation, $Y\rightarrow
1/Y$, and hence the ``transposition relation'' of their respective
scenario-based equality indices. Second, examples \ref{E2} and \ref{E3} can
be obtained via the power transformation $Y\rightarrow Y^{1/\alpha }$ where: the
``original'' random variable $Y$ is governed, respectively, by a gamma law
with density $\psi ( y) =c\exp ( -y) y^{\beta /\alpha
-1}$, and by an inverse-gamma law with density $\psi ( y) =c\exp
( -1/y) y^{-\beta /\alpha -1}$. Third, note that in the case of
example \ref{E4} with identical exponents, $\alpha =\beta $, only the
scenarios $\mathcal{E}=\mathcal{R}$ and $\mathcal{E}=0$ take place.

\emph{A concluding note}: this paper presented a novel Langevin-based method
of scoring the statistical heterogeneity of a wide range of non-negative
size distributions with both finite and infinite means --- be they count,
length, area, volume, mass, energy, duration, etc.; as this paper is
dedicated to the 60$^{\text{th}}$ birthday of Professor Yurij Holovatch, his
students are most welcome to apply this method in the analysis of real-world
data from the scientific fields of interest of Professor Holovatch ---
statistical physics, econophysics, and sociophysics --- as well as from other
fields of science.

\

\ukrainianpart

\title{Досліджуючи нерівності: підхід Ланжевена}
\author{I. Еліазар}
\address{Наукова команда в галузі інновацій розумних пристроїв, Група з розробки нових пристроїв, Корпорація Інтел,
Якум, Ізраїль}

\makeukrtitle

\begin{abstract}

Індекси нерівності є кількісними показниками, які  вимірюють відхилення розподілу багатства в людських суспільствах від ``основного стану''
чистого комунізму. Хоча індекси нерівності були розроблені для застосування в соціоекономіці, вони виявилися ефективно
застосовними в контексті загальних ненегативних розподілів розмірів, таких як кількість, довжина, площа, об'єм, маса, енергія та тривалість.
Індекси нерівності зазвичай базуються на понятті кривих Лоренца, що неявно припускають існування скінчених середніх.
Як наслідок, Лоренц-базовані індекси нерівності  вилучаються з множини нескінчених середніх від розподілів розмірів. В цій статті ми
представляємо індекс нерівності, який базується на  загалом альтернативному підході Ланжевена. Ланжевен-базовані індекси нерівності
вводяться, вивчаються і застосовуються до широкого діапазону ненегативних розподілів розміру як зі скінченими, так і з нескінченими
середніми.

\keywords індекси нерівності, криві Лоренца, рівняння Ланжевена, густина Гіббса, індекс сценарій-базованої рівності
\end{abstract}

\end{document}